\documentclass[11pt]{article}
\usepackage{xspace}
\usepackage{graphicx}
\usepackage{amsmath}
\usepackage{amssymb}
\usepackage{color}
\usepackage{subfigure}  

\definecolor{Red}{rgb}{1,0,0}
\definecolor{Green}{rgb}{0,1,0}
\definecolor{Blue}{rgb}{0,0,1}
\definecolor{Black}{rgb}{0,0,0}

\def\beq{\begin{equation}}
\def\eeq#1{\label{#1}\end{equation}}
\def\eeqn{\end{equation}}

\def\beqa{\begin{eqnarray}}
\def\eeqa#1{\label{#1}\end{eqnarray}}
\def\eeqan{\end{eqnarray}}

\let\bar=\overbar

\def\etal{{\it et al.}}

\def\Dslash{\not{\hbox{\kern-4pt $D$}}}
\def\dslash{\not{\hbox{\kern-2pt $\del$}}}

\def\msb{{\bar{\ssstyle M \kern -1pt S}}}

\def\Title#1{\begin{center} {\Large {\bf #1} } \end{center}}

\begin{document}

\Title{Measurement of the 2$\nu\beta\beta$ decay half-life of $^{150}$Nd to $^{150}$Sm}

\bigskip\bigskip


\begin{raggedright}  

{\it Summer Blot\index{Blot, S.},\\
School of Physics and Astronomy\\
University of Manchester\\
M13 9PL Manchester, UK}\\

\end{raggedright}
\vspace{1.cm}

{\small
\begin{flushleft}
\emph{To appear in the proceedings of the Prospects in Neutrino Physics Conference, 15 -- 17 December, 2014, held at Queen Mary University of London, UK.}
\end{flushleft}
}

\section{Introduction \label{Sec:Introduction}}

The observation of neutrino oscillations has provided evidence for a small, yet finite neutrino mass. The smallness of the neutrino mass is difficult to interpret within the Standard Model (SM), which has prompted the development of a number of Beyond SM theories which attempt to explain the observation. One of the favoured explanations is that the neutrino is fundamentally different from all other SM fermions in that it is a Majorana particle, meaning that the neutrino is its own antiparticle~\cite{Rodejohann:2011mu}. If this is the case, then neutrinoless double beta (0$\nu\beta\beta$) should occur, where the rate of this process is proportional to an effective neutrino mass squared under the assumption that a light Majorana neutrino mediates the decay~\cite{Rodejohann:2011mu}.  In this case, the half-life for the process, $T^{0\nu}_{1/2}$, is expressed via the equation
\small
\begin{equation}
  \label{Eq:halflife_mnu}
  \frac{1}{T^{0\nu}_{1/2}(A,Z)}=
  \vert M^{0\nu}(A,Z)\vert^{2}\cdot G^{0\nu}(Q_{\beta\beta},Z)\cdot\langle m_{\beta\beta}\rangle^{2},
\end{equation}\normalsize
for an isotope with mass number $A$ and atomic number $Z$, where $G^{0\nu}$ is the phase space factor and $M^{0\nu}$ is the nuclear matrix element. The NEMO-3 experiment searches for this rare process in seven different isotopes simultaneously.  The current status of the analysis one of these isotopes, $^{150}$Nd, is reported in these proceedings. This isotope is particularly interesting to use in 0$\nu\beta\beta$ decay search due to its high Q$_{\beta\beta}$ value, 3.37~MeV~\cite{Q_value}, which puts the signal region well above most natural radioactivity. Furthermore, the large Q$_{\beta\beta}$ in combination with its large atomic number gives this isotope the highest G$^{0\nu}$ of all 0$\nu\beta\beta$ decay candidates.

\section{Description of the NEMO-3 experiment \label{Sec:NEMO3_description}}

\begin{figure}[!ht]
\begin{center}
\includegraphics[width=0.5\columnwidth]{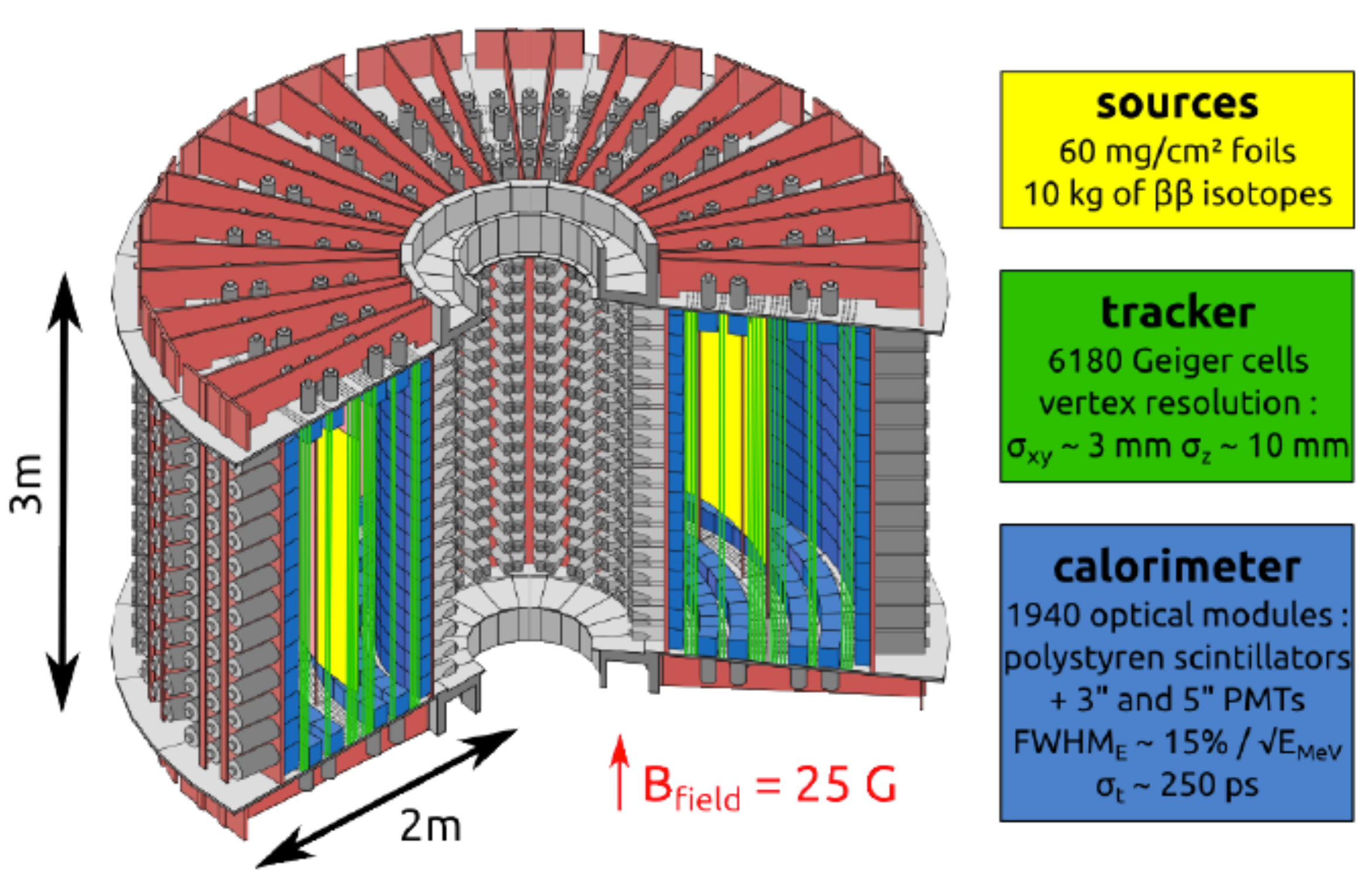}%
\caption{A schematic of the NEMO-3 detector.\label{Fig:detector}}
\end{center}
\end{figure}

The NEMO-3 experiment was located in the Modane Underground Laboratory (LSM), and operated from February 2003 to January 2011.  It is unique among 0$\nu\beta\beta$ decay experiments due to the separation of the candidate isotopes from the active detector system as shown in Figure~\ref{Fig:detector}.  It is cylindrical in shape, with seven different 2$\nu\beta\beta$ decay source foils located at a fixed radius.  The foils are surrounded by two wire tracking chambers, which provide 3D position measurements of charged particles.  Surrounding the tracker on all sides is the calorimeter system, composed of plastic scintillator blocks coupled to low-radioactivity 3'' and 5'' photomultiplier tubes (PMT), which provides both energy and timing measurements. The energy resolution of the calorimeter is $\sigma_{E}/E = 5.8\%\sqrt{E\mathrm{(MeV)}}$ for the optical modules with 5'' PMTs and $\sigma_{E}/E = 7.2\%\sqrt{E\mathrm{(MeV)}}$ for modules with 3'' PMTs. The average time resolution for an electron of 1~MeV is 250~ps.  More information about the detector can be found in~\cite{TDR}.  The separated tracker-calorimeter system allows for direct measurements of background rates, as well as final state kinematics.  In this detector, the experimental signature of a $\beta\beta$ decay event consists of two electrons emitted simultaneously from a common decay vertex.  

The $^{150}$Nd source foil is located in sector 5, where the total mass of the isotope of interest in the foil is 36.55~$\pm$~0.2~g is the isotope of interest~\cite{TDR}.  Two data taking phases are defined for NEMO-3 running. The two phases correspond to the run period before (Phase 1) and after (Phase 2) the instillation of an anti-Radon facility to reduce Rn-induced backgrounds. The combined data from both phases yields a live time of 5.253~yrs, which corresponds to an exposure for $^{150}$Nd of 0.19~kg$\cdot$yrs.

\subsection{The background model}
The $^{150}$Nd source foil contains small amounts of other radioactive isotopes.  The decay rates of these internal backgrounds were measured with a high-purity Germanium (HPGe) detector before the foil was installed in NEMO-3~\cite{TDR}.  The decays of these backgrounds can mimic the $\beta\beta$ decay topology through a number of different processes such as $\beta$ decay followed by  an additional electron produced through M\o ller scattering, Compton scattering of a high energy $\gamma$ ray, or the emission of a conversion electron. Using a number of control channels defined by different final state topologies, the activity of background isotopes, which are both internal and external to the source foil can be directly measured with the NEMO-3 detector, compared with the HPGe measurements when possible, and used to estimate the number of expected events in the 2$\nu\beta\beta$ and 0$\nu\beta\beta$ signal regions.  The most significant backgrounds to the $^{150}$Nd 0$\nu\beta\beta$ decay search are the irreducible 2$\nu\beta\beta$ decay spectrum and the internal contamination of $^{208}$Tl.  Accurate measurements of the decay rates from these two components of the background model are therefore critical for the 0$\nu\beta\beta$ decay search. 

\begin{figure}[t!]
    \subfigure[~$\gamma$ energy, 1e1$\gamma$ channel\label{Fig:1e1g_Eg_log}] {\includegraphics[scale=0.38]{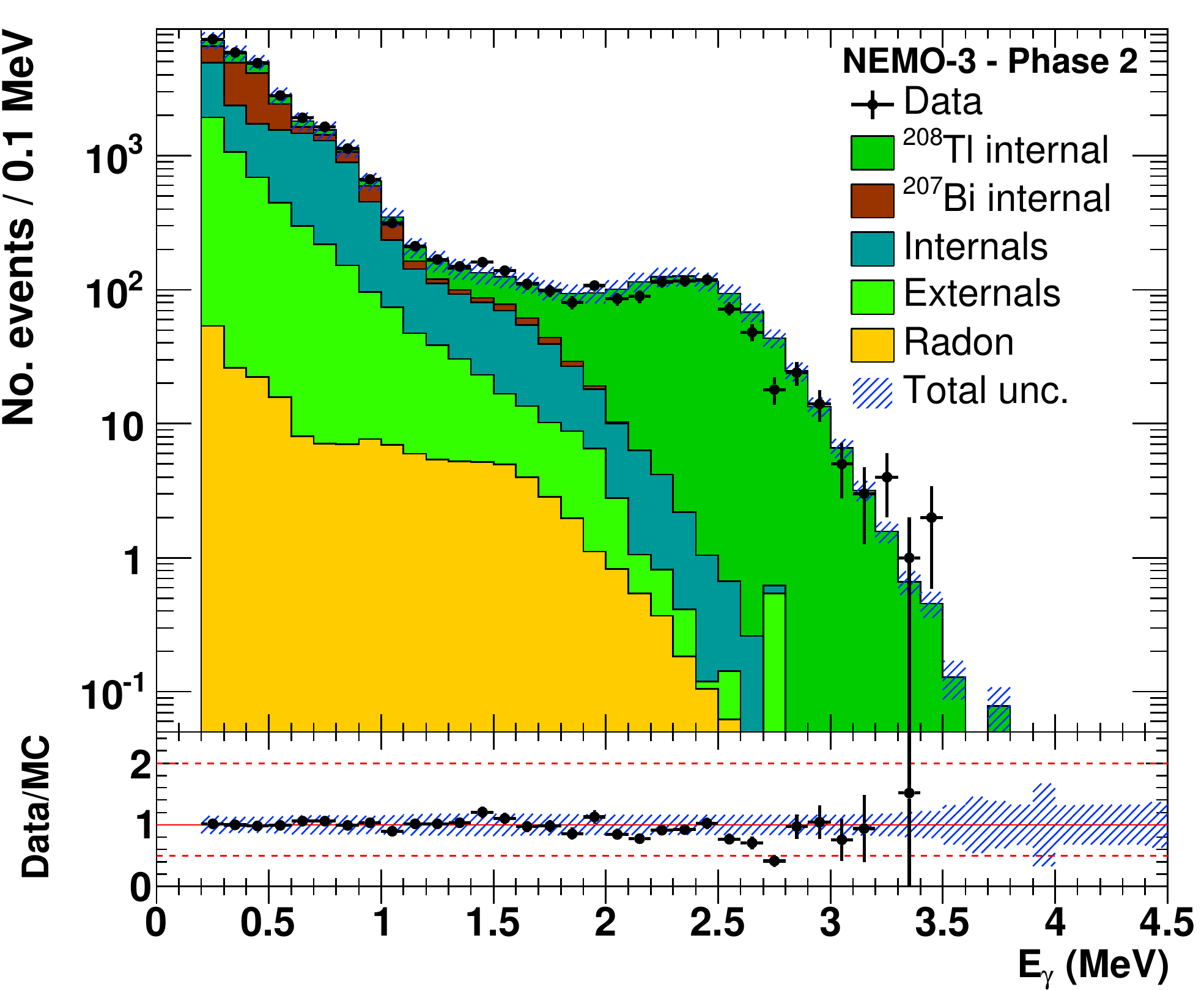}}
    \subfigure[Total energy, 1e2$\gamma$ channel\label{Fig:1e2g_TotalE}] {\includegraphics[scale=0.38]{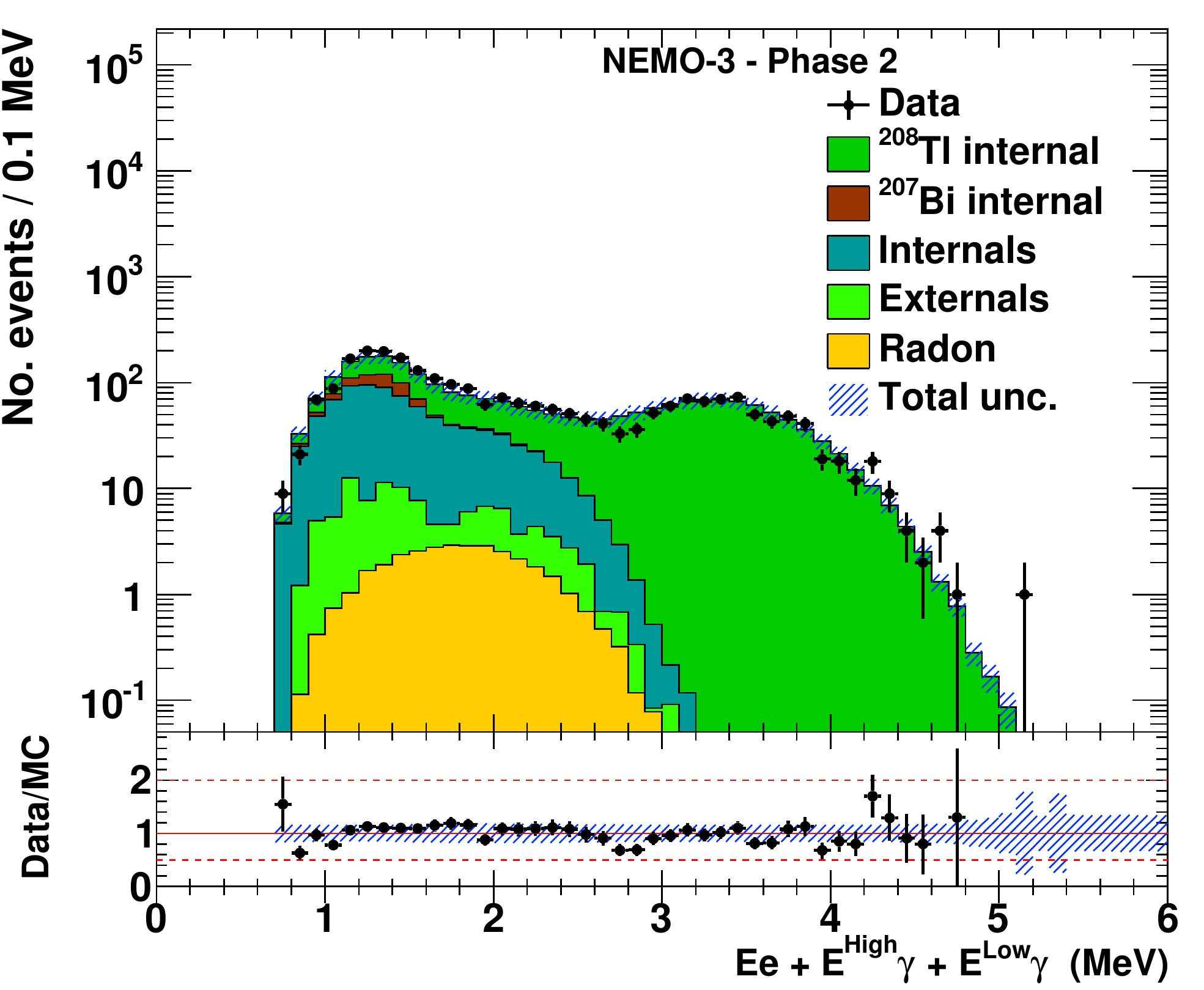}}
    \caption{\label{Fig:1eNg_distributions} The $\gamma$ energy spectrum from the 1e1$\gamma$ channel is shown in Figure~\subref{Fig:1e1g_Eg_log}, and the total energy of all final state particles from the 1e2$\gamma$ channel is shown in Figure~\subref{Fig:1e2g_TotalE}. The data (black points) are compared to simulation (coloured histograms), where the normalization of simulated events reflects the best fit activity for each background source.  The total uncertainty on the model (blue band) includes all statistical and systematic uncertainties and their correlations.}
\end{figure}

The measurements for each background activity as well as the $^{150}$Nd 2$\nu\beta\beta$ decay rate are obtained from a combined likelihood fit to binned observables in both signal and background channels, simultaneously. There are a total of six background channels, which provide sensitivity to various components of the background model.  The activity of internal $^{208}$Tl is mostly constrained by the 1e1$\gamma$ and 1e2$\gamma$ channels, as this isotope primarily $\beta$ decays to the excited state of $^{208}$Pb, which is then followed by high energy $\gamma$ rays from the de-excitation of the daughter nucleus.  The reconstructed $\gamma$ energies from these decays are much higher than most other backgrounds, which provides a nearly background-free region of phase space where the $^{208}$Tl activity can be measured.  Figures~\ref{Fig:1e1g_Eg_log} and~\ref{Fig:1e2g_TotalE} show the binned energy spectra from Phase 2 which are used in the likelihood fit to constrain the $^{208}$Tl activity.  The data is well described by the simulation in both of these figures, particularly at the high energy tails dominated by $^{208}$Tl.

\section{Measurement of the 2$\nu\beta\beta$ half-life}

\begin{figure}[htb!]
    \subfigure[~Total energy\label{Fig:2e_TotalE}] {\includegraphics[scale=0.38]{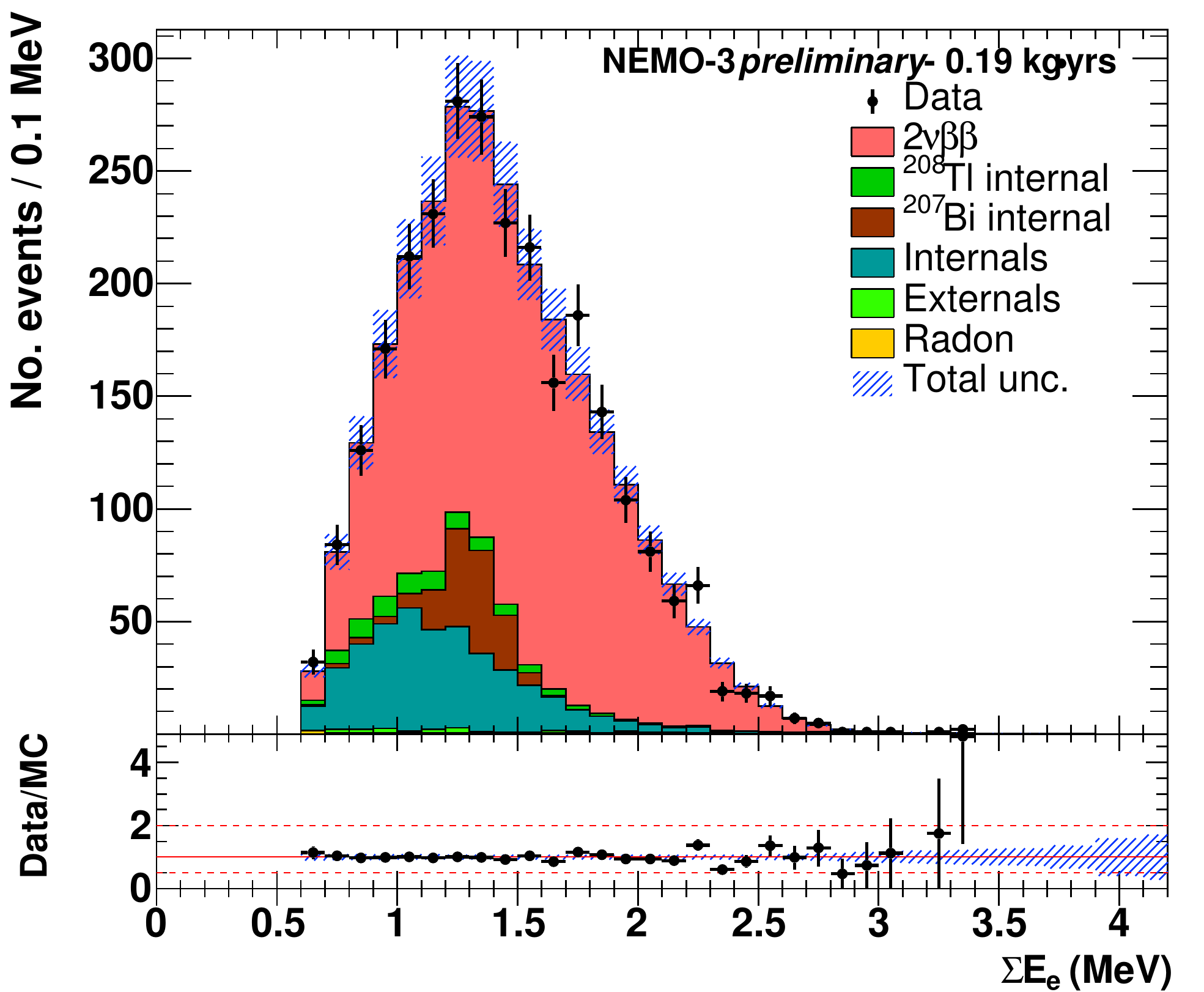}}%
    \subfigure[~Opening angle between tracks\label{Fig:2e_Cosee}] {\includegraphics[scale=0.38]{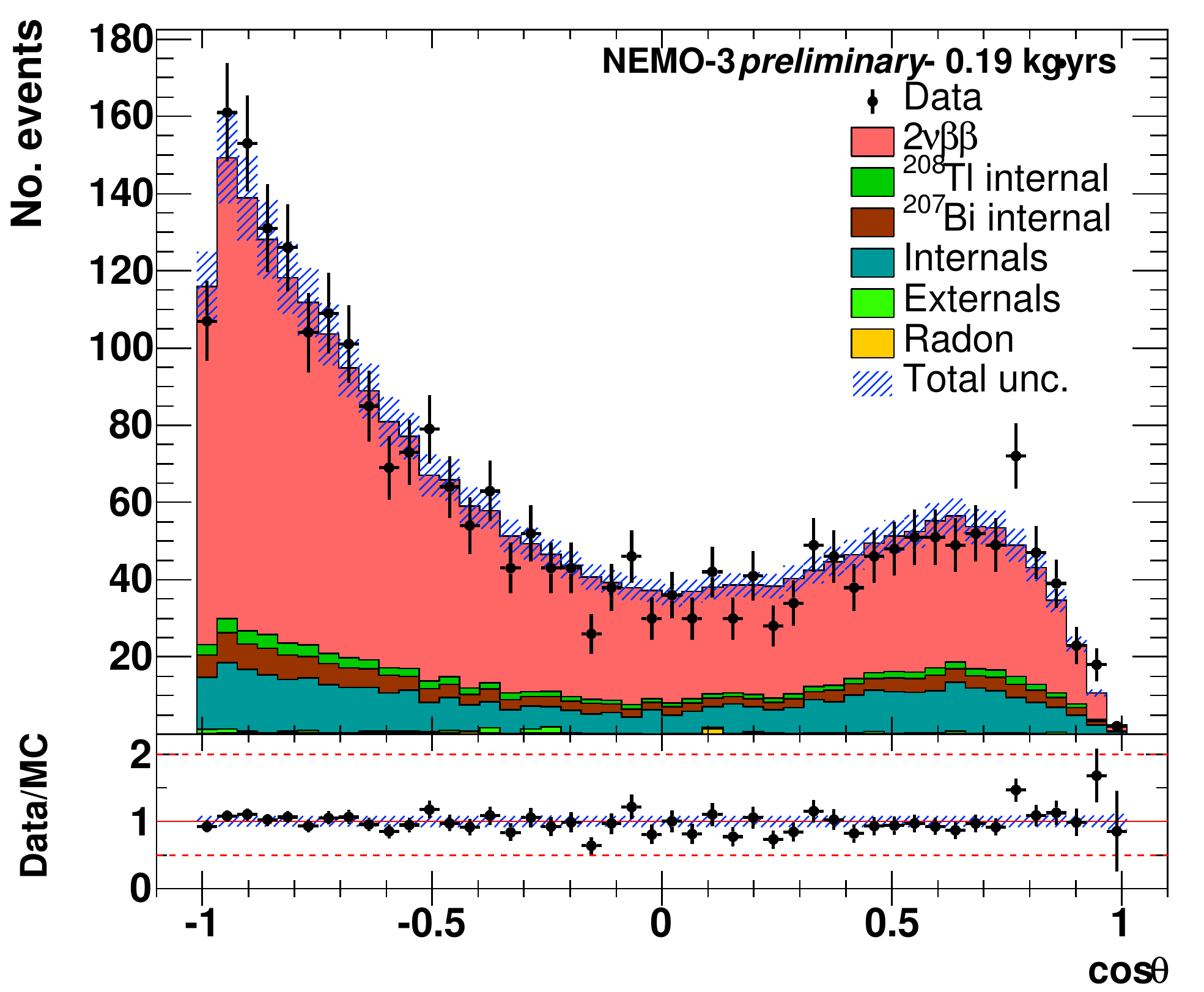}}
   \caption{\label{Fig:2e_channel_observables} 2e channel observables.}
\end{figure}

The two electron channel is used to measure the  2$\nu\beta\beta$ half-life.  Two electron channel candidates must have exactly two tracks with negative curvature and reconstructed vertices inside of the $^{150}$Nd source foil.  The separation between the track vertices on the foil must be $\leq 4$~cm radially and $\leq 8$~cm longitudinally. Each of the tracks must be at least 30~cm long, and associated to separate, isolated scintillator blocks. The impact of the tracks must be on the front face of the scintillator block and the energy of each electron is required to be at least 0.3~MeV, as single electron energies below this threshold are not well described by the simulation. Additional calorimeter hits which do not have a track associated to them are allowed as long the single hit energies are less than 0.2~MeV.  This allows for calorimeter noise in the event which is not simulated. The timing and path lengths of the electrons must be consistent with an internal decay hypothesis.  The total energy of the two electrons and the opening angle between them from events passing these selection criteria are shown in Figures~\ref{Fig:2e_TotalE} and~\ref{Fig:2e_Cosee}, respectively.  With an exposure of 0.19~kg$\cdot$yrs and an efficiency of 4.03\% in the 2e channel, a preliminary 2$\nu\beta\beta$ decay half-life of (9.63 $\pm$ 0.18 (stat.) $\pm$ 0.71 (syst.))$\times 10^{18}$~yrs is obtained with a signal to background ratio 2.88. This is a factor of 1.3 improvement on the statistical error compared to the previously published value for $^{150}$Nd which only used half of the total NEMO-3 exposure~\cite{Argyriades:2008pr}. 

\section{Summary}
The NEMO-3 detector has been used to measure a preliminary 2$\nu\beta\beta$ half-life for $^{150}$Nd of (9.63 $\pm$ 0.18 (stat.) $\pm$ 0.71 (syst.))$\times 10^{18}$~yrs. A likelihood fit to binned observables in both signal and background channels is used to estimate the decay rates of all isotopes considered in the model.  Good agreement between the data and simulation is observed in all signal and background channels. This background model will be validated and refined with future measurements, and used to estimate the number of expected events in the search for the 0$\nu\beta\beta$ decay of $^{150}$Nd.

\end{document}